%\magnification1200
\font\BBig=cmr10 scaled\magstep2
\font\BBBig=cmr10 scaled\magstep3

%%%%%%%%%%%%%%%%%%%%%%%%%%%%%%%%%%%%%%%%%%%
%%%%%%%%%%%%%% the title %%%%%%%%%%%%%%%%%%
%%%%%%%%%%%%%%%%%%%%%%%%%%%%%%%%%%%%%%%%%%%

\def\title{
{\bf\BBBig
\centerline{Chern-Simons gravity based }\bigskip
\centerline{on a non-semisimple group}
}}%%%%% for the front page

\def\foot#1{
\footnote{($^{\the\foo}$)}{#1}\advance\foo by 1
} %%%%% foot(notes
\def\ccr{\cr\noalign{\medskip}}
\def\semidirectproduct{{\ooalign
{\hfil\raise.07ex\hbox{s}\hfil\crcr\mathhexbox20D}}}

%%%%%%%%%%%%%%%%%%%%%%%%%%%%%%%%%%%%%%%%%%%
%%%%%%%%%%%%% the author(s) %%%%%%%%%%%%%%%
%%%%%%%%%%%%%%%%%%%%%%%%%%%%%%%%%%%%%%%%%%%

\def\authors{
\centerline{C.~DUVAL\foot{D\'epartement de Physique,
Universit\'e d'Aix-Marseille~II, and
Centre de Physique
Th\'eorique CNRS--Luminy, Case 907,
F--13288 MARSEILLE Cedex 09 (France);
e-mail: duval@cpt.univ-mrs.fr},
Z.~HORV\'ATH\foot{Institute for Theoretical Physics,
E\"otv\"os University, H--1088 BUDAPEST,
Puskin u.~5-7 (Hungary);
e-mail: zalanh@ludens.elte.hu}
and
P.~A.~HORV\'ATHY\foot{D\'epartement de Math\'ematiques,
Universit\'e de Tours, Parc de Grandmont,
F--37200 TOURS (France);
e-mail: horvathy@univ-tours.fr}}
}

\def\runningauthors{
Duval, Horv\'ath, Horv\'athy
} %%%%% for the header

\def\runningtitle{
Chern-Simons gravity\dots}

%%%%%%%%%%%%%%%%%%%%%%%%%%%%%%%%%%%%%%%%%%%
%%%%%%%%%%%%%% the metrics %%%%%%%%%%%%%%%%
%%%%%%%%%%%%%%%%%%%%%%%%%%%%%%%%%%%%%%%%%%%

%\hsize = 12.5cm %%%%% text width
%\hoffset = 1cm %%%%% hor printing
\voffset = 1cm %%%%% ver printing
\baselineskip = 14pt %%%%% line spacing

\headline ={
\ifnum\pageno=1\hfill
\else\ifodd\pageno\hfil\tenit\runningtitle\hfil\tenrm\folio
\else\tenrm\folio\hfil\tenit\runningauthors\hfil
\fi\fi
} %%%%% header

\nopagenumbers
\footline = {\hfil} %%%%% footer
\pageno=1

%%%%%%%%%%%%%%%%%%%%%%%%%%%%%%%%%%%%%%%%%%%
%%%%%%%%%%%%% greek boldface %%%%%%%%%%%%%%
%%%%%%%%%%%%%%%%%%%%%%%%%%%%%%%%%%%%%%%%%%%

\font\tenb=cmmib10
\newfam\bsfam
 
\textfont\bsfam=\tenb
\mathchardef\betab="080C
\mathchardef\xib="0818
\mathchardef\omegab="0821
\mathchardef\deltab="080E
\mathchardef\epsilonb="080F
\mathchardef\pib="0819
\mathchardef\sigmab="081B
\mathchardef\bfalpha="080B
\mathchardef\bfbeta="080C
\mathchardef\bfgamma="080D
\mathchardef\bfomega="0821
\mathchardef\zetab="0810

%%%%%%%%%%%%%%%%%%%%%%%%%%%%%%%%%%%%%%%%%%%%%%%%%%%%%%
%%%%%%%%%%%%%%%%%% some definitions %%%%%%%%%%%%%%%%%%
%%%%%%%%%%%%%%%%%%%%%%%%%%%%%%%%%%%%%%%%%%%%%%%%%%%%%%

\def\and{\qquad\hbox{and}\qquad}

\def\kikezd{\parag\underbar}

\def\smallcirc{{\raise 0.5pt \hbox{$\scriptstyle\circ$}}}
\def\smallover#1/#2{\hbox{$\textstyle{#1\over#2}$}}
\def\2{{\smallover 1/2}}
\def\ccr{\cr\noalign{\medskip}}
\def\parag{\hfil\break} %%%%% paragraph
\def\={\!=\!}

\def\semidirectproduct{
{\ooalign
{\hfil\raise.07ex\hbox{s}\hfil\crcr\mathhexbox20D}}} %%%% cf. TeX book p.356

%%%%%%%%%%%%%%%%%%%%%%%%%%%%%%%%%%%%%%%%%%%
%%%%%%%%%%%%%%% numberings %%%%%%%%%%%%%%%%
%%%%%%%%%%%%%%%%%%%%%%%%%%%%%%%%%%%%%%%%%%%

\newcount\ch %%%%% ch(apters
\newcount\eq %%%%% eq(uations
\newcount\foo %%%%% foo(tnotes
\newcount\ref %%%%% ref(erences

\def\chapter#1{
\parag\eq = 1\advance\ch by 1{\bf\the\ch.\enskip#1}
}

\def\equation{
\leqno(%\the\ch.
\the\eq)\global\advance\eq by 1
}

\def\reference{
\parag [\number\ref]\ \advance\ref by 1
}

\ch = 1 %%%%% global init ch(apter
%%%%% eq is set to 1 by \chapter
\foo = 1 %%%%% global init foo(tnote
\ref = 1 %%%%% global init ref(erence

%%%%%%%%%%%%%%%%%%%%%%%%%%%%%%%%%%%%%%%%%%%
%%%%%%%%%%%%%%%% the text %%%%%%%%%%%%%%%%%
%%%%%%%%%%%%%%%%%%%%%%%%%%%%%%%%%%%%%%%%%%%

\title
\vskip .5cm
\authors
\vskip .08in
\parag
{\bf Abstract.}
{\sl The gauge theory-formulation of string-motivated lineal gravity
proposed by Cangemi and Jackiw
is obtained by dimensional reduction from a 
$(2+1)$ dimensional gravity with Chern-Simons Lagrangian.
}
\bigskip

%\chapter{Introduction}

The `string-inspired' gravity in $(1+1)$ dimensions [1] admits a
gauge-theory formulation [2]: starting with the centrally extended
Poincar\'e algebra with $4$ generators,
$$
[P_a,J]=\epsilon_a^bP_b,
\qquad
[P_a,P_b]=\epsilon_{ab}i\Lambda I
\equation
$$
($a,\,b=1,2$), where $I$ is the central element, the `Poincar\'e' gauge theory
with connection form
$$
A=e^aP_a+\omega J+a\2i\Lambda I
\equation
$$
and Lagrangian
$$
{\cal L}=\eta_AF^A=\eta_a(De)^a+\eta_2d\omega+
\eta_3(da+\2e^a\epsilon_{ab}e^b),
\qquad
\eta_A=(\eta_a,\eta_2,\eta_3),
\equation
$$
where the $\eta$'s are Lagrange multipliers,
leads to the $(1+1)$ dimensional gravitational equation [1]
$$
R=0
\and
\partial_\mu\partial_\nu\eta=\2\Lambda h_{\mu\nu}
\equation
$$
(where $h_{\mu\nu}={\rm diag}(1,-1)$ and 
$\Lambda$ is the cosmological constant), supplemented with the additional equation
$$
da+\2e^a\epsilon_{ab}e^b=0.
\equation
$$

The algebra (1) has been subsequently extensively studied in connection of
a recently proposed Wess-Zumino-Witten model [3-6]. 

In the conslusion of their paper [2],
Cangemi and Jackiw wonder whether their model can be obtained from
a $(2+1)$ dimensional model by reduction. The aim of this Letter is to 
show that this is indeed possible. Using recent results on
non-semisimple algebras [6,7], we construct a $(2+1)$-dimensional
Chern-Simons Lagrangian whose reduction gives precisely the 
Cangemi-Jackiw model. Other models are found in Ref. [8-9].

%\chapter{The Extended Poincar\'e algebra in $2+1$ dimensions}

Consider the extended Poincar\'e algebra in $2+1$ dimensions
with non-vanishing commutators
$$\matrix{
[J_\mu,J_\nu]=\epsilon_{\mu\nu}^\rho J_\rho,
\qquad\hfill
&[J_\mu,P_\nu]=\epsilon_{\mu\nu}^\rho P_\rho,\hfill
\ccr
[J_\mu,T_\nu]=\epsilon_{\mu\nu}^\rho T_\rho,\hfill
&[P_\mu,P_\nu]=\epsilon_{\mu\nu}^\rho T_\rho,
\cr}
\equation
$$
($\mu,\, \nu,\, \rho=0,1,2$).
Note that the $T_\mu$'s commute with the $P_\mu$'s and with 
each other. In these formulae, the indices are raised and lowered through the
Lorentz metric $\eta_{\mu\nu}={\rm diag}(-1,1,1)$.
This algebra carries a symmetric, non-degenerate bilinear form,
namely
$$
\Omega_{AB}M^AM^B=
P_\mu P^\mu+J_\mu T^\mu+T_\mu J^\mu.
\equation
$$
(The other Casimir invariant, $T_\mu T^\nu$ is degenerate since the group 
is non semi-simple). The group generated by the algebra (1.1) can be 
represented by upper triangular matrices [10]
$$
\pmatrix{
U&U\omega&U(\2\omega^2+a)\ccr
&U&U\omega\ccr
&&U%\cr
},
%\equation
%$$
%with$$
\qquad\left\{
\eqalign{
&U=\pmatrix{u&v\cr v^*&u^*\cr},
\cr
&a=\pmatrix{a_2&i(a_0-a_1)\cr
i(a_0+a_1)&-a_2\cr},
\cr
&\omega=\pmatrix{\omega_2&i(\omega_0-\omega_1)\cr
i(\omega_0+\omega_1)&-\omega_2\cr}.
\cr}\right.
\equation
$$
Note that $U\in {\rm SU}(1,1)$. %,\, \omega\in\;? a\in\; ?$.

%\chapter{Gauge theory on our algebra}

Consider the gauge theory in $2+1$ dimensions
given by the connection form
$$
A_\alpha=
e_\alpha^\mu P_\mu+\omega_\alpha^\mu J_\mu+a_\alpha^\mu T_\mu,
\equation
$$
where $\mu=0,1,2$ is an algebra index and $\alpha=0,1,2$ denote
space-time indices. Its curvature form is found as
$$\eqalign{
F_{\alpha\beta}=&\big(
\partial_\alpha e_\beta^\rho
-\partial_\beta e_\alpha^\rho
+\epsilon_{\mu\nu}^\rho\omega_\alpha^\mu e_\beta^\nu
-\epsilon_{\mu\nu}^\rho\omega_\beta^\mu e_\alpha^\nu\big)P_\rho
\ccr
&+\big(\partial_\alpha\omega_\beta^\rho
-\partial_\beta\omega_\alpha^\rho
+\epsilon_{\mu\nu}^\rho\omega_\alpha^\mu\omega_\beta^\nu\big)J_\rho
\ccr
&+\big(\partial_\alpha a_\beta^\rho+
\epsilon_{\mu\nu}^\rho\omega_\alpha^\mu a_\beta^\nu
-\partial_\beta a_\alpha^\rho
-\epsilon_{\mu\nu}^\rho\omega_\beta^\mu a_\alpha^\nu
+\epsilon_{\mu\nu}^\rho e_\alpha^\mu e_\beta^\nu
\big)T_\rho.
\cr}
\equation
$$
The first bracket here is the torsion and the second bracket is
the curvature of the spin connection.

Now we want to postulate a Chern-Simons gauge theory. The Chern-Simons 
$3$-form is conveniently found by considering the closed $4$-form
$$
(F,*F)=\epsilon^{\alpha\beta\gamma\delta}
\Omega_{AB}F^A_{\alpha\beta} F^B_{\gamma\delta}
\equation
$$
which is locally the divergence of the Chern-Simons $3$-form,
$\partial_\alpha
({\rm Chern-Simons})^\alpha$. 
Our Chern-Simons gauge theory is therefore given by the Lagrangian
$$
{\cal L}=\2\epsilon^{\alpha\beta\gamma}
\left[\big(\partial_\alpha e_\beta^\mu e_\gamma^\nu
+\partial_\alpha\omega_\beta^\mu a_\gamma^\nu
+\partial_\alpha a_\beta^\mu\omega_\gamma^\nu\big)\eta_{\mu\nu}
+\epsilon_{\mu\nu\rho}\big(
\omega_\alpha^\mu e_\beta^\nu e_\gamma^\rho+
\omega_\alpha^\mu\omega_\beta^\nu a_\gamma^\rho\big)
\right].
\equation
$$

The corresponding field equations read
$$\eqalign{
&\epsilon^{\alpha\beta\gamma}\big(
\partial_\alpha e_\beta^\nu
+\epsilon_{\mu\rho}^\nu\omega_\alpha^\mu e_\beta^\rho
\big)=0,
\ccr
&\epsilon^{\alpha\beta\gamma}\big(
\partial_\alpha\omega_\beta^\nu
+\2\epsilon_{\mu\rho}^\nu\omega_\alpha^\mu\omega_\beta^\rho
\big)=0,
\ccr
&\epsilon^{\alpha\beta\gamma}\big(
\partial_\alpha a_\beta^\nu
+\2\epsilon_{\mu\rho}^\nu e_\alpha^\mu e_\beta^\rho
+\epsilon_{\mu\rho}^\nu\omega_\alpha^\mu a_\beta^\rho
\big)=0.
\cr}
\equation
$$
The first two equations say that the torsion and the curvature 
has to vanish.

%\chapter{Dimensional Reduction}

Dimensional reduction of our theory reproduces
the formalism of Cangemi and Jackiw [2]. Let us in fact assume that 
none of the
fields depends on the second coordinate $x_2$. Using the indices 
$\mu=(k,2)$ and $\alpha=(a,2)$,
$$
\matrix{
e_b^k\to e_b^k,
\qquad\hfill
&e_2^k\to\eta^k,\hfill
\ccr
a_2^2\to\eta^2,\hfill
&\omega_2^2\to\eta^3,
\ccr
\omega_a^2\to\omega_a,\hfill
&a_a^2\to a_a,
\ccr
a_\alpha^k=\omega_\alpha^k=0.\hfill
&\cr}
\equation
$$
Then the Chern-Simons Lagrangian (12) reduces precisely to the Lagrangian 
(1.3) of Cangemi and Jackiw.

\kikezd{Note added}: 
This paper has been written in 1993, but remained unpublished because of its overlap with
Cangemi's Salamanca talk [11].
The structure studied here  has been applied to various physical problems and in particular to string theory [3,12].
Similar conclusions to those presented here were reached later, independently,
by de Montigny at al [13].

\vskip7mm
%%%%%%%%%%%%%%%%%%%%%%%%%%%%%%%%%%%%%%%%%%%
%%%%%%%%%%%%% the references %%%%%%%%%%%%%%
%%%%%%%%%%%%%%%%%%%%%%%%%%%%%%%%%%%%%%%%%%%

\centerline{\bf\BBig References}

\reference
H. Verlinde, in {\it The Sixth Marcel Grossmann Meeting on
General Relativity}, edited by M. Sato 
(World Scientific, Singapore, 1992);
C. Callan, S. Giddings, A. Harvey and A. Strominger,
Phys. Rev. {\bf D45}, 1005 (1992).

\reference
D. Cangemi and R. Jackiw,
Phys. Rev. Lett. {\bf 69}, 233 (1992);
Phys. Lett. {\bf B299}, 24 (1993);
Ann. Phys. {\bf 225}, 229 (1993).

\reference
A. Ach\'ucarro and P. K. Townsend,
Phys. Lett. {\bf B180}, 89 (1986);
E. Witten, Nucl. Phys. {\bf B311}, 47 (1988/89).

\reference
C. Nappi and E. Witten, Phys. Rev. Lett. {\bf 71}, 3751 (1993).

\reference
E. Witten,
Commun. Math. Phys. {\bf 117}, 353 (1988).

\reference
A. Medina and Ph. Revoy, Ann. Sci. \'Ec. Norm. Sup. 
{\bf 18}, 553 (1985);
J. M. Figueroa-O'Farrill and S. Stanciu,
hep-th/9402035

\reference
D. I. Olive, E. Rabinovici and A. Schwimmer,
Phys. Lett. {\bf 321B}, 355 (1994)

\reference
C. Klim\v cik and A. A. Tseytlin, Phys. Lett. {\bf B323}, 305 (1994);
K. Steftsos, Phys. Rev. {\bf D50}, 2784 (1994)
N. Mohammedi, Phys. Lett. {\bf B331}, 93 (1994);
E. Kiritsis and C. Kounnas Phys. Lett. {\bf B320}, 264 (1994)

\reference
A. Ach\'ucarro,
Phys. Rev. Lett. {\bf 70}, 1037 (1993);
A. Ach\'ucarro and M. E. Ortiz,
Phys. Rev. {\bf D48}, 3600 (1993).

\reference
N. Ja. Vilenkin and A. U. Klimyk,
{\it Representation of Lie Groups and Special Functions}. Recent Advances (Mathematics and Its Applications.
(1994).

\reference
D. Cangemi, Phys. Lett. {\bf B297}, 261 (1992);
in Proc. {\it Group Theoretical Methods in Phys.}, Salamanca'92,
del Olmo, Santander, Guilarte (eds.), Anales de Fisica Monografias,
Vol. II, p. 72 (1993). See also
Cangemi, Phys. Lett. {\bf B297}, 261 (1992).

\reference
pp waves were first discussed by
H. W. Brinkmann,
Math. Ann. {\bf 94}, 119-145 (1925), and
have been rediscovered and applied to strings in
C. Duval, G. Gibbons and P. A. Horv\'athy, 
Phys. Rev. {\bf  D43}, 3907 (1991). See
A. A. Tseytlin, Nucl. Phys. {\bf B390}, 153 (1993);  
C. Duval, Z. Horv\'ath and P. A. Horv\'athy, 
Phys. Lett. {\bf B313}, 10 (1993);
Mod. Phys. Lett. {\bf A8}, 3749 (1993). 

\reference
M. de Montigny, F. C. Khanna,
A. E. Santana, E. S. Santos and J. D. M. Vianna,
% Poincar\'e gauge theory and Galilean covariance
Annals Phys. {\bf 277} 144 (1999).

%%%%%%%%%%%
\bye